\documentclass[preprint,eqsecnum]{revtex4}
\usepackage{amsmath}
\usepackage{graphicx}
\newcommand{\ds}{\displaystyle}
\begin{document}
\title{Football fever: goal distributions and non-Gaussian statistics}
\author{Elmar~Bittner}
\author{Andreas~Nu{\ss}baumer}
\author{Wolfhard~Janke}
\affiliation{Institut f\"ur Theoretische Physik and Centre for 
Theoretical Sciences (NTZ) -- Universit\"at Leipzig,\\
Augustusplatz 10/11, D-04109 Leipzig, Germany}

\author{Martin~Weigel}
\affiliation{Department of Mathematics, 
  Heriot-Watt University, Riccarton, Edinburgh,
  EH14\,4AS, Scotland, UK}
\date{\today}

\begin{abstract}
  
  Analyzing football score data with statistical techniques, we investigate how the
  not purely random, but highly co-operative nature of the game is reflected in
  averaged properties such as the probability distributions of scored goals for the
  home and away teams. As it turns out, especially the tails of the distributions are
  {\em not\/} well described by the Poissonian or binomial model resulting from the
  assumption of uncorrelated random events. Instead, a good effective description of
  the data is provided by less basic distributions such as the negative binomial one
  or the probability densities of extreme value statistics. To understand this
  behavior from a microscopical point of view, however, no waiting time problem or
  extremal process need be invoked. Instead, modifying the Bernoulli random process
  underlying the Poissonian model to include a simple component of {\em
    self-affirmation\/} seems to describe the data surprisingly well and allows to
  understand the observed deviation from Gaussian statistics. The phenomenological
  distributions used before can be understood as special cases within this framework.
  We analyzed historical football score data from many leagues in Europe as well as
  from international tournaments, including data from all past tournaments of the ``FIFA World
  Cup'' series, and found the proposed models to be applicable rather universally.  In
  particular, here we analyse the results of the German women's premier football league 
  and consider the two separate German men's premier leagues in the East
  and West during the cold war times and the unified league after 1990 to see how
  scoring in football and the component of self-affirmation depend on cultural and
  political circumstances.

\end{abstract}

\maketitle

\section{Introduction}

Football is perhaps the most popular sports in Europe, attracting millions of
spectators and involving thousands of players each year.  As a traditional
socio-cultural institution of significant economical importance, football has also
been the subject of numerous scientific efforts, for instance geared towards the
improvement of game tactics, the understanding of the social effects of the fan
scene etc.  Much less effort has been devoted, it seems, to the understanding of
football (and other ball sports) from the perspective of the stochastic behavior of
co-operative ``agents'' (i.e., players) in abstract models. This problem as well as
many other topics relating to the statistical properties of socially interacting
systems have recently been identified as fields where the model-based point-of-view
and methodological machinery of statistical mechanics might add a new perspective to
the much more detailed investigations of more specific disciplines
\cite{stauffer:03}.

Score distributions of football and other ball games have been occasionally
considered by mathematical statisticians for more than fifty years
\cite{moroney:56,reep:71,pollard:73,clarke:95,dyte:00,malacarne:00,green}. Initially,
the limited available data were found to be reasonably modeled by the Poissonian
distribution resulting from the simplest assumption of a completely random process
with a fixed (but possibly team dependent) scoring probability \cite{moroney:56}. In
the following, it was empirically found that a better fit could be produced with a
negative binomial distribution originally introduced as an {\em ad hoc\/} measure of
generalizing the parameter range for fitting certain biological data
\cite{arbous:51}. The negative binomial form occurs naturally for a mixture of Poissonian
processes with a certain distribution of (independent) success probabilities
\cite{reep:71}.  Furthermore, recently it was found \cite{green} that score
distributions of some football leagues are better described by the
generalized distributions of extreme value statistics \cite{book}, while others
rather follow the negative binomial distribution. This yielded a rather inhomogeneous
picture and, more generally, for a system of highly co-operative entities it might be
presumed that models without correlations cannot be an adequate description. What is
more, all these proposals remained in the realm of observation, since the considered
statistical models where selected by best fit, without offering any microscopical
justification for the choice.

The distribution of extremes, i.e., the probability density function of
($k^\mathrm{th}$) maximal or minimal values of independent realizations of a random
variable, is described by only a few universality classes, depending on the asymptotic
behavior of the original distribution \cite{book}. Apart from the direct importance
of the problem of extremes in actuarial mathematics and engineering, generalized
extreme value (GEV) distributions have been found to occur in such diverse systems as
the statistical mechanics of regular and disordered systems
\cite{boucaud,bramwell:98,bramwell:00,berg,dayal,bittner}, turbulence
\cite{noullez:02} or earth quake data \cite{varotsos:05}. However, in most cases
global properties were considered instead of explicit extremes, and the occurrence of
GEV distributions led to speculations about hidden extremal processes in these
systems, which could not be identified in most cases, though. It was only realized
recently that GEV distributions can also arise naturally as the statistics of sums of
{\em correlated\/} random variables \cite{dahlstedt:01,bertin:05,bertin}, which could
explain their ubiquity in physical systems.

For the problem of scoring in football, correlations naturally occur through
processes of (positive and negative) feedback of scoring on both teams, and we shall
see how the introduction of simple rules for the adaptation of the success
probabilities in a modified Bernoulli process upon scoring a goal leads to systematic
deviations from Gaussian statistics. We find simple models with a single parameter of
{\em self-affirmation\/} to best describe the available data, including cases with
relatively poor fits of the negative binomial distribution. The latter is shown to
result from one of these models in a particular limit, explaining the relatively good
fits observed before. For the models under consideration, exact recurrence relations
and precise closed-form approximations of the probability density functions can be
derived.  Although the limiting distributions of the considered models in general do
{\em not\/} follow the statistics of extremes, it is demonstrated how alternative
models leading to GEV distributions could be constructed. The best fits are found for
models where each extra goal encourages a team even more than the previous one: a
true sign of {\em football fever\/}.

The rest of the paper is organized as follows. Section \ref{sec:probability}
discusses the probability distributions used by us and previous authors to fit
football score data and their relations to the microscopic models introduced here.
The results of fits of the considered models and distributions to the data are
summarized and discussed in Sec.\ \ref{sec:data} with emphasis on a comparison of the
goal distributions in the divided Germany of the cold war times and of the German 
women's and men's premier leagues, and an analysis of
the results of the ``FIFA World Cup'' series. Finally, Sec.\ \ref{sec:concl} contains our
conclusions, and some of the statistical technicalities of the considered modified
binomial models are summarized in App.~\ref{sec:appendix}.

\section{\label{sec:probability}Probability distributions and
microscopic models}

The most obvious and readily available global property characterizing a football
match is certainly given by the overall score of the game. Hence, to investigate the
balance of chance and skill in football \cite{reep:71}, here we consider the
distributions of goals scored by the home and away teams in football league or cup
matches. To the simplest possible approximation, both teams have independent and
constant probabilities of scoring during each appropriate time interval of the match, thus degrading
football to a pure game of chance. Since the scoring probabilities will be
small, the resulting probabilities of final scores will follow a Poissonian
distribution,
\begin{gather}
    \label{eq:poisson}
    P^\mathrm{h}_{\lambda_{h}}(n_\mathrm{h}) = 
      \frac{{\lambda_\mathrm{h}}^{n_\mathrm{h}}}{n_\mathrm{h}!} \exp(-\lambda_\mathrm{h}),\;\;\;
    P^\mathrm{a}_{\lambda_{a}}(n_\mathrm{a}) = 
      \frac{{\lambda_\mathrm{a}}^{n_\mathrm{a}}}{n_\mathrm{a}!} \exp(-\lambda_\mathrm{a}),
\end{gather}
where $n_h$ and $n_a$ are the final scores of the home and away teams, respectively,
and the parameters $\lambda_\mathrm{h}$ and $\lambda_\mathrm{a}$ are related to the
average number of goals scored by a team, $\lambda=\langle n \rangle$.
As an additional check of the fit to the data, one might then also
consider the probability densities of the sum $\sigma = n_\mathrm{h}+n_\mathrm{a}$
and difference $\delta = n_\mathrm{h}-n_\mathrm{a}$ of goals scored,
\begin{equation}
  \label{eq:poisson_sum_diff}
  \begin{array}{rcl}
    \ds P^\Sigma_{\lambda_\mathrm{h},\lambda_\mathrm{a}}(\sigma) & = &
    \ds\sum_{n=0}^\sigma P^\mathrm{h}_{\lambda_\mathrm{h}}(n) P^\mathrm{a}_{\lambda_\mathrm{a}}(\sigma-n) = 
    \frac{(\lambda_\mathrm{h}+\lambda_\mathrm{a})^\sigma}{\sigma!}
    \exp[-(\lambda_\mathrm{h}+\lambda_\mathrm{a})], \\[2ex]
    \ds P^\Delta_{\lambda_\mathrm{h},\lambda_\mathrm{a}}(\delta) & = & 
    \ds \sum_{n=0}^\infty P^\mathrm{h}_{\lambda_\mathrm{h}}(n+\delta) P^\mathrm{a}_{\lambda_\mathrm{a}}(n) = 
    e^{-(\lambda_\mathrm{h}+\lambda_\mathrm{a})} 
    \left(\frac{\lambda_\mathrm{h}}{\lambda_\mathrm{a}}\right)^{\delta/2} 
    I_\delta (2 \sqrt{\lambda_\mathrm{h} \lambda_\mathrm{a}}),
  \end{array}
\end{equation}
where $I_\delta$ is the modified Bessel function (see \cite{abramowitz:70}, p.~374).
Note that $P^\Sigma_{\lambda_\mathrm{h},\lambda_\mathrm{a}}(\sigma)$ is itself a
Poissonian distribution with parameter $\lambda =
\lambda_\mathrm{h}+\lambda_\mathrm{a}$.

Clearly, the assumption of constant and independent scoring probabilities for the
teams is not appropriate for real-world football matches. Since we are interested in
averages over the matches during one or several seasons of a football league or cup,
one might expect a {\em distribution\/} of scoring probabilities $\lambda$ depending
on the different skills of the teams, the lineup for the match, tactics, weather
conditions etc., leading to the notion of a {\em compound\/} Poisson distribution. It
can be easily shown \cite{fisz,feller} that for the special case of the scoring
probabilities $\lambda$ following a gamma distribution,
\begin{equation}
  \label{eq:gamma}
f(\lambda) = \left\{
  \begin{array}{l@{\hspace{0.5cm}}l}
    \displaystyle \frac{a^r}{\Gamma(r)}\lambda^{r-1}\mathrm{e}^{-a\lambda},& \lambda > 0,\\
    0, & \lambda \le 0,
  \end{array}
\right.
\end{equation}
the resulting compound Poisson distribution has the form of a negative binomial
distribution (NBD),
\begin{equation}
  \label{eq:neg_binomi}
   P_{r,p}(n) = \int_0^\infty\mathrm{d}\lambda\,P_\lambda(n) f(\lambda) =\frac{\Gamma(r+n)}{n!\,\Gamma(r)} p^n(1-p)^r,
\end{equation}
where $p = 1/(1+a)$. The negative binomial form has been found to describe football
score data rather well \cite{pollard:73,green}. The underlying assumption of the
scoring probabilities following a gamma distribution seems to be rather {\em ad
  hoc\/}, however, and fitting different seasons of our data with the Poissonian
model \eqref{eq:poisson}, the resulting distribution of the parameters $\lambda$ does
not resemble the gamma form \eqref{eq:gamma}. Analogous to
Eq.~\eqref{eq:poisson_sum_diff}, for the negative binomial distribution~\eqref{eq:neg_binomi}
one can evaluate the
probabilities for the sum $\sigma$ and difference $\delta$ of goals scored by the
home and away teams,
\newcommand{\ph}{{p_\mathrm{h}}}
\newcommand{\pa}{{p_\mathrm{a}}}
\newcommand{\rh}{{r_\mathrm{h}}}
\newcommand{\ra}{{r_\mathrm{a}}}
\begin{equation}
  \label{eq:nbd_sum_diff}
  \begin{array}{rcl}
    \ds P^\Sigma_{\ph,\rh,\pa,\ra}(\sigma)
    & =  &\ds(1-\ph)^\rh (1-\pa)^\ra \pa^\sigma \frac{ \Gamma(\ra + \sigma)}{\sigma!\,
      \Gamma(\ra)}\,\null_2F_1\left(-\sigma, \rh; 1 - \sigma - \ra;
      \frac{\ph}{\pa}\right), \\[2ex]
    \ds P^\Delta_{\ph,\rh,\pa,\ra}(\delta) & = & 
    \ds (1-\ph)^\rh (1-\pa)^\ra \ph^\delta \frac{\Gamma(\rh + \delta)}{
        \delta!\,\Gamma(\rh)}\,
    \null_2 F_1\left(\rh + \delta, \ra; 1 + \delta; \ph \pa\right),
  \end{array}
\end{equation}
where $\null_2 F_1$ is the hypergeometric function  (see
\cite{abramowitz:70}, p.~555). Restricting $\ph=\pa$, the distribution
of the total score simplifies to $P^\Sigma_{p,r,q,s}(\sigma) =
P_{p,r+s}(\sigma)$, i.e., one finds a composition law
similar to the case of the Poissonian distribution.

To do justice to the fact that playing football is different from playing dice, one
has to take into account that goals are not simply independent events but, instead,
scoring certainly has a profound feedback on the motivation and possibility of
subsequent scoring of both teams (via direct motivation/demotivation of the players,
but also, e.g., by a strengthening of defensive play in case of a lead), i.e., there
is a fundamental component of (positive or negative) feedback in the system. We do so
by introducing such a feedback effect into the bimodal model (being the discrete
version of the Poissonian model \eqref{eq:poisson} above): consider a football match
divided into $N$ time steps (we restrict ourselves here to the natural choice $N =
90$, but good fits are found for any choice of $N$ within reasonable limits) with
both teams having the possibility to either score or not score in each time step.
Feedback is introduced into the system by having the scoring probabilities $p$ depend
on the number $n$ of goals scored so far, $p=p(n)$. Several possibilities arise. For
our model ``A'', upon each goal the scoring probability is modified as
\begin{equation}
  p(n) = p(n-1) + \kappa,
\end{equation}
with some fixed constant $\kappa$ (unless $p(n-1) + \kappa > 1$, in which case $p(n)
= 1$, or $p(n-1) + \kappa < 0$, which is replaced by $p(n) = 0$). Alternatively, one
might consider a multiplicative modification rule,
\begin{equation}
  p(n) = \kappa p(n-1) 
\end{equation}
(again modified to ensure $0\le p(n)\le 1$), which we refer to as model ``B''. The
resulting modified binomial distributions $P_N(n)$ for the total number of goals
scored by one team can be computed exactly from a Pascal type recurrence relation,
\begin{equation}
  \label{eq:recurrence}
  P_N(n) = [1-p(n)] P_{N-1}(n) + p(n-1)P_{N-1}(n-1),
\end{equation}
where, e.g., $p(n) = p_0+\kappa n$ for model ``A'' and $p(n) = p_0\kappa^n$ for model
``B''. 
Eq.~(\ref{eq:recurrence}) is intuitively plausible, since $n$ successes in $N$ trials can
be reached either from $n$ successes in $N-1$ trials plus a final failure or from
$n-1$ successes in $N-1$ trials and a final success.
For a more formal proof see the discussion in App.\ \ref{sec:appendix}, where 
for the additive case of
model ``A'', it is also demonstrated that the continuum limit
of $P_N(n)$, i.e., $N\rightarrow \infty$ with $p_0 N$ and $\kappa N$ kept fixed, is given
by the negative binomial distribution \eqref{eq:neg_binomi} with $r = p_0/\kappa$ and
$p = 1 - \mathrm{e}^{-\kappa N}$ (note that this also includes the ``generalized
binomial distribution'' considered in Refs.~\cite{drezner:93,drezner:06}). Thus the
good fit of a negative binomial distribution to the data can be understood from the
``microscopic'' effect of self-affirmation of the teams or players, without making
reference to the somewhat poorly motivated composition of the pure Poissonian model with
a gamma distribution.  
Finally,
the assumption of independence of the scoring of the home and away teams can be
relaxed by coupling the adaptation rules upon scoring, for instance as
\begin{equation}
  \label{eq:modelC}
  \begin{array}{rcl@{\hspace{0.5cm}}rcl@{\hspace{0.5cm}}l}
    p_\mathrm{h}(n) &=& p_\mathrm{h}(n-1) \kappa_\mathrm{h}, &
    p_\mathrm{a}(n) &=& p_\mathrm{a}(n-1) / \kappa_\mathrm{a}, &
    \text{for a goal of the home team},\\
    p_\mathrm{h}(n) &=& p_\mathrm{h}(n-1) / \kappa_\mathrm{h}, &
    p_\mathrm{a}(n) &=& p_\mathrm{a}(n-1) \kappa_\mathrm{a}, &
    \text{for a goal of the away team}, 
  \end{array}
\end{equation}
which we refer to as model ``C''. If both teams have $\kappa>1$, this results in an
incentive for the scoring team and a demotivation for the opponent. But a value
$\kappa <1$ is conceivable as well. The probability density function
$P_N(n_\mathrm{h}, n_\mathrm{a})$ can be computed recursively as well, cf.\
App.~\ref{sec:appendix}.

Starting from the observation that the goal distributions of certain leagues do not
seem to be well fitted by the negative binomial distribution, Greenhough {\em et
  al.\/} \cite{green} considered fits of the GEV distributions,
\begin{equation}
  \label{eq:gev}
  \begin{array}{rcl@{\hspace{0.3cm}}l}
    P_{\xi,\mu,\sigma}(n) & = &
    \displaystyle
    \frac{1}{\sigma}\left(1+\xi\frac{n-\mu}{\sigma}\right)^{-1-1/\xi}
    \exp\left[-\left(1+\xi\frac{n-\mu}{\sigma}\right)^{-1/\xi}\right]
    & \text{for} \quad \xi\neq 0,\\
    P_{\mu,\sigma}(n) & = &
    \displaystyle
    \frac{1}{\sigma}\exp\left[-\exp\left(-\frac{n-\mu}{\sigma}\right)
      -\frac{n-\mu}{\sigma}\right]
    & \text{for} \quad \xi=0,
  \end{array}
\end{equation}
to the data, obtaining good fits in some cases. According to the value of the
parameter $\xi$, these distributions are known as Weibull ($\xi < 0$), Gumbel ($\xi =
0$) and Fr\'echet ($\xi > 0$) distributions, respectively. As for the case of the
negative binomial form as a compound Poisson distribution, the use of extremal value
statistics appears here rather {\em ad hoc\/}. We would like to point out, however,
that the GEV distributions indeed can result from a modified microscopical model with
feedback. To this end, consider again a series of trials for a number $N$ of time
steps. Assume that the probability to score $U_1$ goals in time step 1 is distributed
according to $P_1(U_1)=P(U_1)$ (e.g., with a Poisson distribution $P$), the
probability to score $U_2$ goals in time step 2 is $P_2(U_2)=P(U_1+U_2)/Z_2$ etc.,
such that $P_i(U_i) = P(\sum_{j=1}^{i-1} U_j + U_i)/Z_i$. For any continuous
distribution $P$, this means that due to the normalization factors $Z_i$ the
distribution of $U_i$ will have enhanced tails compared to the distribution of
$U_{i-1}$ (unless $U_{i-1} = 0$) etc., resulting in a positive feedback effect
similar to that of models ``A'', ``B'' and ``C''. We refer to this prescription as
model ``D''. From the results of Bertin and Clusel \cite{bertin:05,bertin} it then follows that
the limiting distribution of the total score $n = \sum_{i=1}^N U_i$ is a GEV
distribution, where the specific form of distribution [in particular the value of the
parameter $\xi$ in \eqref{eq:gev}] depends on the falloff of the original
distribution $P$ in its tails.


\section{\label{sec:data}Data and results} 

Concerning football matches played in leagues, our main data set consists of matches
played in Germany, namely for the ``Bundesliga'' (men's premier league FRG, 1963/64
-- 2004/05, $\approx 12\,800$ matches), the ``Oberliga'' (men's premier league GDR,
1949/50 -- 1990/91, $\approx 7700$ matches), and for the ``Frauen-Bundesliga''
(women's premier league FRG, 1997/98 -- 2004/05, $\approx 1050$ matches)
\cite{daten1,daten2,daten3,daten4}. Our focus was here to see how in particular the
feedback effects reflected in the football score distributions depend on cultural and
political circumstances and are possibly different between men's and women's leagues.
We first determined histograms estimating the probability density functions (PDFs)
$P^\mathrm{h}(n_h)$ and $P^\mathrm{a}(n_a)$ of the final scores of the home and away
teams, respectively~\cite{fn1}.
Similarly, we determined histograms for the PDFs
$P^\Sigma(\sigma)$ and $P^\Delta(\delta)$ of the sums and differences of final
scores. To arrive at error estimates on the histogram bins, we utilized the bootstrap
resampling scheme \cite{bootstrap}.

\begin{figure}
\centerline{
\includegraphics[scale=0.90]{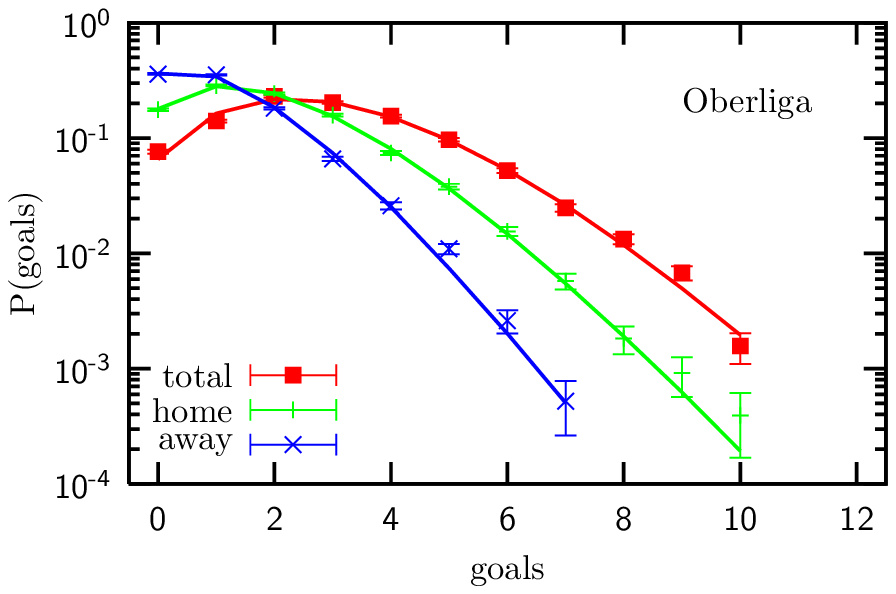}
\includegraphics[scale=0.90]{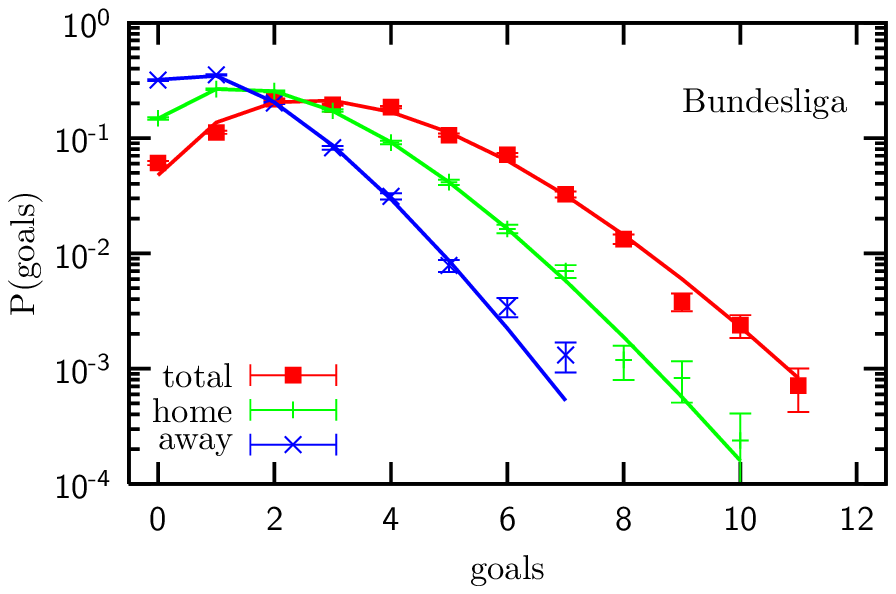}
}
\caption{Probability density of goals scored by home teams, away teams, and of the
  total number of goals scored in the match.
Left: ``Oberliga'' of the GDR between 1949 and 1990.   
Right: ``Bundesliga'' of the FRG in the seasons of 1963/64 -- 1990/91.   
The lines for ``home'' and ``away'' show fits of the negative binomial distribution
(\ref{eq:neg_binomi}) to the data; the line for ``total'' denotes the resulting
distribution of the sum, Eq.~(\ref{eq:nbd_sum_diff}).
}
\label{fig_all}
\end{figure}

\begin{figure}
\centerline{
\includegraphics[scale=0.90]{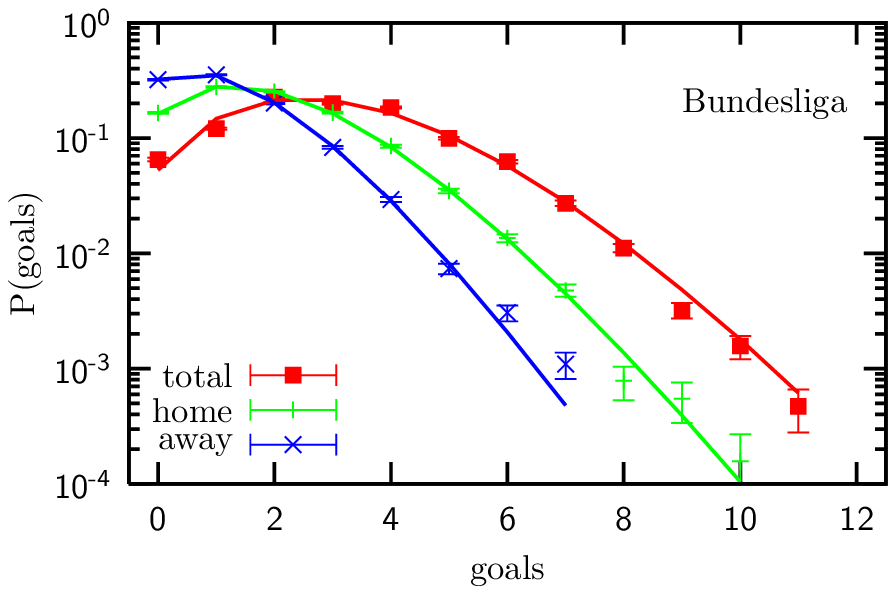}
\includegraphics[scale=0.90]{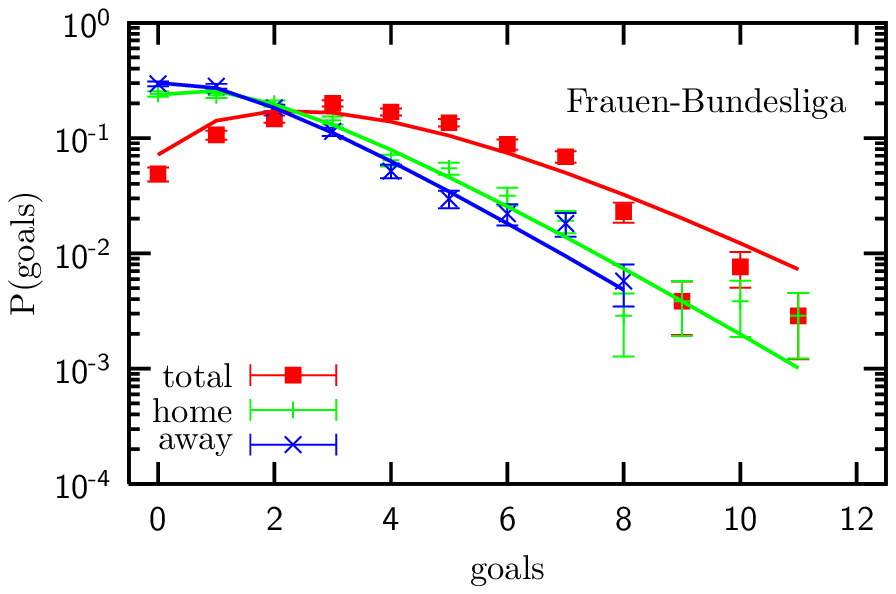}
}
\caption{Probability density of goals scored in the German premier league ``Bundesliga'' for all seasons (left) and
in the women's ``Frauen-Bundesliga'' (right).}
\label{fig_1991}
\end{figure}

We first considered fits of the PDFs of the phenomenological descriptions considered
previously, namely the Poissonian form (\ref{eq:poisson}), the negative binomial
distribution (\ref{eq:neg_binomi}) and the distributions (\ref{eq:gev}) of extreme
value statistics. The parameters of fits of these types to the data are summarized in
Table~\ref{tab_ddr_dfb} comparing the East German ``Oberliga'' to the West German
``Bundesliga'' (1963/64 -- 1990/91, $\approx 8400$ matches) during the time of the
German division, and in Table~\ref{tab_frauen} comparing the data for all games of
the German men's premier league ``Bundesliga'' to the German women's
premier league ``Frauen-Bundesliga''. Not to our surprise, and in accordance with previous
findings \cite{reep:71,green}, the simple Poissonian ansatz (\ref{eq:poisson}) is not
found to be an adequate description for any of the data sets. Deviations occur here
mainly in the tails with large numbers of goals which in general are found to be
fatter than can be accommodated by a Poissonian model, whereas the distribution peaks
are reasonably well represented. On the contrary, the negative binomial form
(\ref{eq:neg_binomi}) models all of the considered data well as is illustrated with
fits of the corresponding form to our data in Fig.~\ref{fig_all} comparing
``Oberliga'' and ``Bundesliga'' and in Fig.~\ref{fig_1991} presenting ``Bundesliga''
and ``Frauen-Bundesliga''. Comparing the leagues, we find that the parameters $r$ of the NBD
fits for the ``Bundesliga'' are about twice as large as for the ``Oberliga'', whereas
the parameters $p$ are smaller for the ``Bundesliga'', cf.\ the data in
Table~\ref{tab_ddr_dfb}. Recalling that the form (\ref{eq:neg_binomi}) is in fact the
continuum limit of the feedback model ``A'' discussed above, these differences
translate into larger values of $\kappa$ and smaller values of $p_0$ for the
``Oberliga'' results. That is to say, scoring a goal in a match of the East German
premier league was a more encouraging event than scoring a goal in a match of the
West German league. Alternatively, this observation might be interpreted as a
stronger tendency of the perhaps more professionalized teams of the West German league to
switch to a strongly defensive mode of play in case of a lead.  Consequently, the
tails of the distributions are slightly fatter for the ``Oberliga'' than for the
``Bundesliga''. Comparing the results for the ``Frauen-Bundesliga'' to those for the
``Bundesliga'', even more pronounced tails are found for the former, resulting in
very significantly larger values of the self-affirmation parameter $\kappa$ for the
matches of the women's league, see the fit parameters collected in
Table~\ref{tab_frauen} and the fits of the NBD type presented in Fig.~\ref{fig_1991}.

Considering the fits of the GEV distributions (\ref{eq:gev}) to the data for all
three leagues, we find that extreme value statistics are in general a reasonably good
description of the data. The shape parameter $\xi$ is always found to be small in
modulus and negative in the majority of the cases, indicating a distribution of the
Weibull type (which is in agreement with the findings of Ref.~\cite{green}). On the
other hand, fixing $\xi = 0$ yields overall clearly larger values of $\chi^2$ per
degree-of-freedom, indicating that the data are hardly compatible with a distribution
of the Gumbel type. Comparing ``Oberliga'' and ``Bundesliga'', we consistently find
larger values of the parameter $\xi$ for the former, indicative of the comparatively
fatter tails of these data discussed above, see the data in Table~\ref{tab_ddr_dfb}.
The location parameter $\mu$, on the other hand, is larger for the West German league
which features a larger average number of goals per match (which can be read off also
more directly from the $\lambda$ parameter of the Poissonian fits), 
while the scale parameter $\sigma$ is similar for both leagues. Comparing to the results 
for the NBD, we do not find any cases where the GEV distributions would
provide the best fit to the data, so clearly the leagues considered here are not of
the type of the general ``domestic'' league data for which Greenhough {\em et al.\/}~\cite{green}
found better matches with the GEV than for the NBD statistics. 
Similar conclusions hold true for the comparisons of ``Bundesliga'' and
``Frauen-Bundesliga'', with the latter taking on the role of the ``Oberliga''.

Assuming, for the time being, that the histograms of the final scores of the home and
away teams are properly modeled by the fits presented in Tables~\ref{tab_ddr_dfb} and
\ref{tab_frauen}, it is worthwhile as a consistency check to see whether the
resulting estimates (\ref{eq:poisson_sum_diff}) and (\ref{eq:nbd_sum_diff}) of the
PDFs for the Poisson and negative binomial distributions are consistent with the data
for the sums and differences. Of course, such consistency can only be expected if the
histograms of home and away scores are statistically independent, which assumption
certainly is a strongly simplifying approximation. In Table~\ref{tab_fits} we
summarize the mean squared deviations $\chi^2$ of the PDFs
(\ref{eq:poisson_sum_diff}) resp.\ (\ref{eq:nbd_sum_diff}), evaluated with the
parameters of the fits to the home and away scores of Tables~\ref{tab_ddr_dfb} and
\ref{tab_frauen}, from the data for the sums and differences. While again clearly the
Poissonian ansatz disqualifies as an acceptable model of the data, the NBD fits the
data for the ``Oberliga'' and the ``Frauen-Bundesliga'' comparatively well, cf.\ the data in
Table~\ref{tab_fits} and the ``total'' fits in Figs.~\ref{fig_all} and \ref{fig_1991}.
For the ``Bundesliga'', however, significant deviations are observed. These
deviations might go back to an effect of correlation between the home and
away scores.
To investigate this question we computed the empirical correlation coefficient,
\begin{equation}
  \label{eq:correlation_coefficient}
  R = \frac{\mathrm{Cov}(n_\mathrm{h}, n_\mathrm{a})}{\sigma(n_\mathrm{h})\sigma(n_\mathrm{a})},
\end{equation}
where $\sigma(n)$ denotes the square root of the variance of $n$ and
$\mathrm{Cov}(n_\mathrm{h}, n_\mathrm{a})$  the covariance of $n_\mathrm{h}$ and
$n_\mathrm{a}$. We find $R =-0.015 \pm  0.011$ for the ``Oberliga'' and $R = -0.031 \pm  0.009$ for the
``Bundesliga'', indicating stronger home-away score correlations for the 
``Bundesliga''~\cite{fn2}.

\begin{figure}
\centerline{
\includegraphics[scale=0.90]{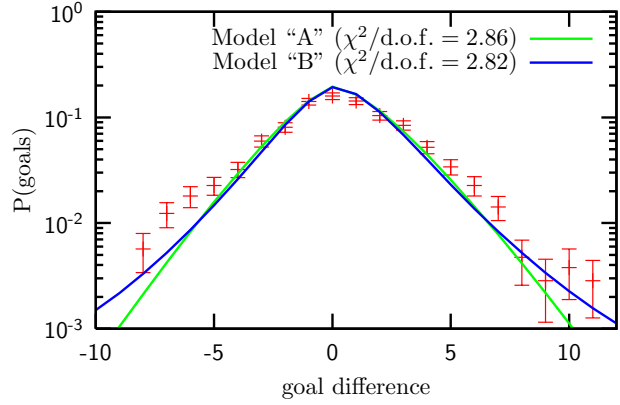}
}
\caption{Goal differences in the German women's premier league together with fits of
models ``A'' and ``B''.}
\label{fig:diff}
\end{figure}

In total, the best fits so far are clearly achieved by the NBD ansatz. Since this
distribution is obtained only as the continuum limit of the microscopic model ``A'',
it is interesting to see how fits of the {\em exact\/} distribution (for $N=90$) 
resulting from the recurrence (\ref{eq:recurrence}) for model ``A'', but also fits of the
multiplicatively modified binomial distribution of model ``B'' compare to the results
found above. We perform fits to the exact distributions of both models by employing
the simplex method \cite{num_rec} to minimize the total $\chi^2$ of the data for the
home and away scores. Alternatively, we also considered fitting additionally to the
sums and differences in a simultaneous fit and found very similar results with an
only slight improvement of the fit quality for the sums and differences at the
expense of somewhat worse fits for the home and away scores. We summarize the fit
results in Table~\ref{tab_meanner}. We also performed fits to the more elaborate
model ``C'', but found rather similar results to the simpler model ``B'' and hence do
not present the results here. Comparing the results of model ``A'' to the fits of the
limiting NBD, we find almost identical fit qualities for the final scores of both
teams. However, the sums and differences of scores are considerably better described
by model ``A'', indicating that here the deviations from the continuum limit are
still relevant. In Fig.~\ref{fig:diff}, we present the
differences of goals in the German women's premier league together with the fits of
models ``A'' and ``B''. The multiplicative model ``B'', where each goal motivates a
team even more than the previous one, within the statistical errors yields fits of
the same quality as model ``A'', such that a distinct advantage cannot be attributed
to either of them, cf.\ the data in Table~\ref{tab_meanner}.

\begin{figure}
\centerline{
\includegraphics[scale=0.90]{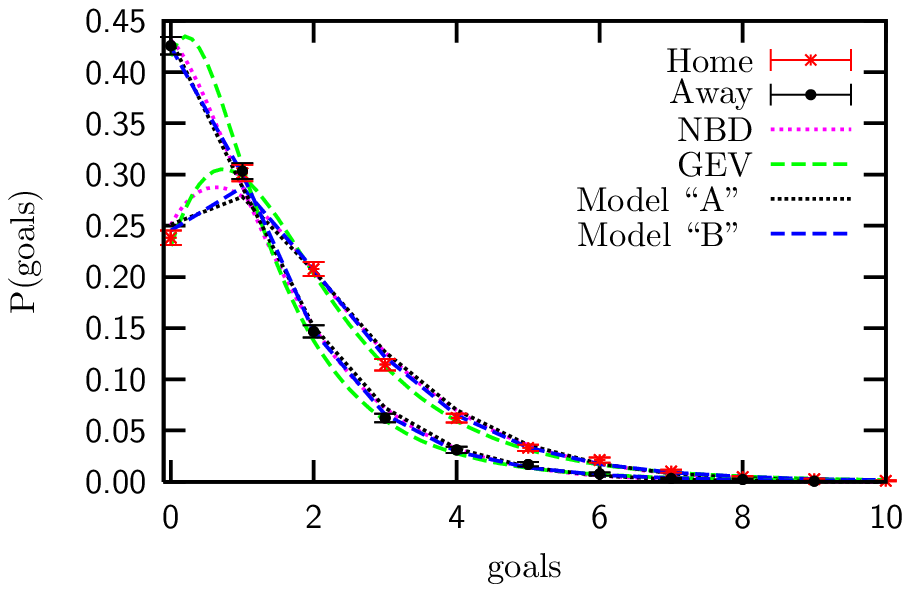}
\includegraphics[scale=0.90]{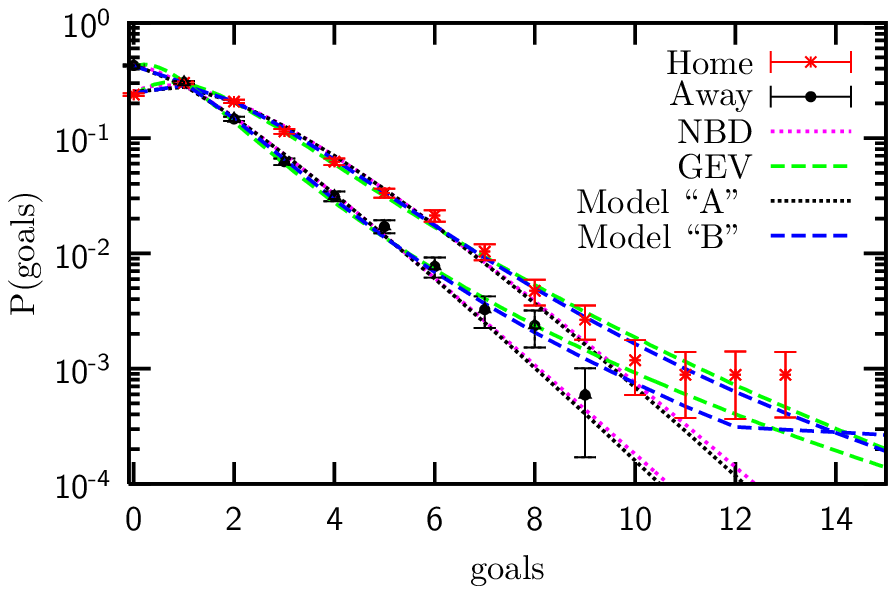}
}
\caption{Probability density of goals scored by the home and away teams 
in the qualification stage of the ``FIFA World Cup'' series from 1930 to 2002.
}
\label{fig:wm}
\end{figure}

Finally, to leave the realm of German football, we considered the score data of the
``FIFA World Cup'' series from 1930 to 2002, focusing on the results from the
qualification stage ($\approx 3400$ matches) \cite{daten5}~\cite{fn3}.
The results of fits of the
phenomenological distributions (\ref{eq:poisson}), (\ref{eq:neg_binomi}) and
(\ref{eq:gev}) as well as the models ``A'' and ``B'' are collected in
Table~\ref{tab_wm}. Compared to the domestic league data discussed above, the results
of the World Cup show distinctly heavier tails, cf.\ the presentation of the data in
Fig.~\ref{fig:wm}. Considering the fit results, this leads to good fits for the
heavy-tailed distributions, and, in particular, in this case the GEV distribution
provides a better fit than the negative binomial model, similar to what was found by
Greenhough {\em et al.\/}~\cite{green} for some of their data. This difference to the
German league data discussed above can be attributed to the possibly very large
differences in skill between the opposing teams occurring since all countries are
allowed to participate in the qualification round. A glance back to 
Table~\ref{tab_frauen} reveals a remarkable similarity with the parameters of
the ``Frauen-Bundesliga'' (e.g., in both cases the NBD parameters $p$ are comparatively large 
while $r$ is small, and the GEV parameters $\xi$ are positive), where a similar
explanation appears quite plausible since the very good players are concentrated
in two or three teams only.
Turning to the fits of the models
``A'' and ``B'', we again find model ``A'' to fit rather similar to its continuum
approximation, the NBD. On the other hand, model ``B'' describes the data extremely well, 
for the away team even better than the GEV distributions (\ref{eq:gev}). It is, of course, also
possible and interesting to analyze the results from the final round. Similar to
other cups such as the German ``DFB-Pokal'' we also considered, the rules are slightly
different here, since no game can end in a draw, leading to special correlation
effects in particular in the histograms of the goal differences. These problems will
be investigated in a forthcoming publication.

\section{\label{sec:concl}Summary} 

We have considered German domestic and international football score data with respect
to certain phenomenological probability distributions as well as microscopically
motivated models. The Poisson distribution resulting from the assumption of 
independent scoring probabilities for the opposing teams does not provide a
satisfactory fit to any of our data. Many data sets are rather well described by the
negative binomial distribution considered before \cite{reep:71}, however, some cases
have heavier tails than can be accommodated by this distribution and, instead, rather
follow a distribution from extreme value statistics.

We have shown that football score data can be understood from a certain class of
modified binomial models with a built-in effect of self-affirmation of the teams upon
scoring a goal. The negative binomial distribution fitting many of the data sets can
in fact be understood as a limiting distribution of our model ``A'' with an additive
update rule of the scoring probability. It is found, however, that the {\em exact\/}
distribution of model ``A'' provides in general rather better fits to the data than
the limiting NBD, in particular concerning the sums and differences of goals scored.
However, it does not provide very good fits in cases with
heavier tails such as the qualification round of the ``FIFA World Cup'' series. The
variant model ``B'', on the other hand, where a multiplicative update rule ensures
that each goal motivates the team even more than the previous one, fits these
world-cup data as well as the data from the German domestic leagues extremely well.
Thus, the contradicting evidence for better fits of some football score data
with negative binomial and other data with GEV distributions is reconciled with the
use of a plausible microscopic model covering both cases. We also analyzed results
from further leagues, such as the 
Austrian, Belgian, British, Bulgarian, Czechoslovak, Dutch, French, Hungarian,
Italian, Portuguese, Romanian, Russian, Scottish and Spanish
premier leagues, and arrived at similar conclusions.

Comparing the score data between the separate German premier leagues during the cold
war times, we find heavier tails for the East German league. In terms of our
microscopic models, this corresponds to a stronger component of self-affirmation as
compared to the West German league. Similarly, the German women's premier league
``Frauen-Bundesliga'' shows a much stronger feedback effect than the men's premier
league, with at first sight surprisingly many parallels to the ``FIFA World Cup'' series. 
In general, we find less professionalized leagues to feature stronger
components of positive feedback upon scoring a goal, perhaps indicating a still
stronger infection with the football fever there \dots

It is obvious that the presented models with a single parameter of self-affirmation
are a gross over-simplification of the complex psycho-social phenomena on a football
pitch. It is all the more surprising then, how rather well they model the considered
score distributions~\cite{fn4}. 
Naturally,
however, a plethora of opportunities for improvement of the description and further
studies opens up. For instance, considering averages over whole leagues or cups, we
have not taken into account the differences in skill between the teams. Likewise, if
time-resolved scoring data were made available, a closer investigation of the
intra-team and inter-team motivation and demotivation effects would provide an
intriguing future enterprise to undertake.


\begin{acknowledgments}
  The authors are grateful to O.\ Penrose and S.\ Zachary for fruitful discussions.
  This work was partially supported by the Deutsche Forschungsgemeinschaft (DFG)
  under grant No.\ JA483/22-1 and the EU RTN-Network `ENRAGE': {\em Random Geometry
  and Random Matrices: From Quantum Gravity to Econophysics\/} under grant
  No.~MRTN-CT-2004-005616.
  M.W.\ acknowledges support by the EC ``Marie Curie
  Individual Intra-European Fellowships'' programme under contract No.\
  MEIF-CT-2004-501422.
\end{acknowledgments}

\appendix

\section{\label{sec:appendix}Probabilistics of correlated Bernoulli trials}

Consider a series of $N$ Bernoulli random variables $U_i$, $i=1,\ldots,N$, with
probabilities $1-p_i$ and $p_i$ for the outcomes ``0'' (``failure'') and ``1''
(``success''), respectively. We are interested in the distribution $P_N(\sum_{i=1}^N
U_i = n)$ of the number of successes in $N$ trials. For the limiting case of equal
and constant probabilities $p_i = p$, $i=1,\ldots,N$, the $U_i$ are i.i.d.\ random
variables and $P_N$ is given by the binomial distribution
\begin{equation}
  \label{eq:binomi}
  P_N(\sum_{i=1}^N U_i = n) = {N \choose n} p^n\,(1-p)^{N-n},
\end{equation}
which is a properly normalized (discrete) probability distribution function according
to the binomial theorem. This can be generalized for arbitrary independent choices of
probabilities $p_i$.

We discuss a more general case where, instead, the probabilities $p_i$ themselves
depend on the number of previous successes, $p_i = p(\sum_{k=1}^{i-1} U_i)$. Due to
the introduced correlations, one should then consider the joint probability
distribution of the $U_i$,
\begin{equation}
  \label{eq:joint}
  P(U_1,\ldots,U_N) = \prod_{i=1}^N \left\{p(\sum_{k=1}^{i-1} U_k)\delta_{U_i,1}+
    [1-p(\sum_{k=1}^{i-1} U_k)]\delta_{U_i,0}\right\},
\end{equation}
from which the desired distribution of successes follows as the marginal
$P_N(n) = \sum_{\{U_i\}}P(U_1,\ldots,U_N) \delta_{\sum U_i, n}$. Instead of formally
proceeding from (\ref{eq:joint}) it is more convenient, however, to observe that the
{\em distances\/} $D_j$ between subsequent successes are independent geometrically
distributed random variables with probabilities $1-p(n)$, i.e.\ $P(D_j = d_j) =
p(j)[1-p(j)]^{d_j-1}$, $j=0,\ldots,n$, and the desired marginal distribution becomes
\begin{eqnarray}
  P_N(n) & = & \sum_{d_0=1}^{N-n} \cdots \sum_{d_n=1}^{N-n}\,
  [1-p(0)]^{d_0-1}p(0)\cdots p(n-1)[1-p(n)]^{d_n-1} \delta_{\sum_j d_j, N}\nonumber\\
  & = & \prod_{j=0}^{n-1}p(j) \sum_{d_0=1}^{N-n} \cdots \sum_{d_n=1}^{N-n}\,
  \prod_{j=0}^n[1-p(j)]^{d_j-1}\,\delta_{\sum_j d_j, N}.
  \label{eq:marginal}
\end{eqnarray}

Manipulating this form it is straightforward to prove a Pascal
type recurrence relation for the probabilities $P_N(n)$,
\begin{equation}
  \label{eq:pascal}
  P_N(n) = [1-p(n)] P_{N-1}(n) + p(n-1)P_{N-1}(n-1),
\end{equation}
which together with the initial condition $P_0(0) = 1$ and noting that $P_N(n) = 0$
for $n>N$ allows to construct the distribution with an O($N^2$) computational effort
compared to the formal O($2^N$) effort implied by Eq.~(\ref{eq:joint}).
Multiplying (\ref{eq:pascal}) by
$u^n$ and summing over all $n$, one arrives at
\begin{equation}
  \label{eq:difference_eq}
  G_N(u)-G_{N-1}(u) = (u-1)H_{N-1}(u),
\end{equation}
where
\begin{equation}
  \label{eq:generating_function}
  G_N(u) = \sum_{n=0}^\infty P_N(n) u^n,\;\;\;H_N(u) = \sum_{n=0}^\infty p(n) P_N(n) u^n,
\end{equation}
such that $G_N(u)$ is the generating function of $P_N(n)$. The continuum limit $N
\mapsto t$ is thus described by the differential equation 
\begin{equation}
  \label{eq:differential_eq}
  \frac{\partial G(u,t)}{\partial t} = (u-1)H(u,t).
\end{equation}

The additive, correlated binomial model discussed in the main text modifies $p
\mapsto p + \kappa$ on each success, unless $p + \kappa > 1$ in which case $p \mapsto
1$. Restricting ourselves to the range of parameters where $p < 1$, we have $p(n) =
p_0 + \kappa n$, $H_N(u) = p_0G_N(u) + \kappa u \frac{\partial}{\partial u}G_N(u)$
and Eq.~(\ref{eq:differential_eq}) becomes
\begin{equation}
  \label{eq:binomi_diff}
  \frac{\partial G(u,t)}{\partial t} = (u-1)[p_0 G(u,t) + \kappa u
  \frac{\partial}{\partial u}G(u,t)],
\end{equation}
which is readily checked to be solved by
\begin{equation}
  \label{eq:neg_binomi_gen}
  G(u,t) = [\mathrm{e}^{\kappa t}-u(\mathrm{e}^{\kappa t}-1)]^{-p_0/\kappa}.
\end{equation}
Hence, $P(n)$ has a negative binomial distribution \cite{arbous:51,fisz},
\begin{equation}
  \label{eq:neg_binomi_app}
  P_t(n) = \mathrm{e}^{-p_0 t}
  \frac{\Gamma(p_0/\kappa+n)}{n!\,\Gamma(p_0/\kappa)}{\left(1-\mathrm{e}^{-\kappa t}\right)}^n
  = {r+n-1 \choose n} p^n(1-p)^r,
\end{equation}
where $r = p_0/\kappa$ and $p =1- \mathrm{e}^{-\kappa t}$. For $N\kappa =
\mathrm{const} < 1$, this continuum approximation is appropriate in the same limit
where the Poissonian distribution is a valid approximation for the binomial
distribution (\ref{eq:binomi}), i.e., for $N \gg 1$ with $N p_o =
\mathrm{const}$.

For the multiplicative, correlated binomial model, after each success the probability
is modified as $p \mapsto \kappa p$ (unless $\kappa p > 1$, in which case $p \mapsto
1$), such that $p(j) = p_0 \kappa^j$ for the range of parameters where $p(n) < 1$. In
this case, the differential equation (\ref{eq:differential_eq}) becomes
\begin{equation}
  \label{eq:multi_differential}
  \frac{\partial G(u,t)}{\partial t} = (u-1)p_0 G(\kappa u,t).
\end{equation}
Note that due to the different first arguments of $G$, this is {\em not\/} an ordinary
differential equation. We currently do not see how the solution could be expressed in
terms of elementary or special functions in this case.
 Still, the distribution
$P_N(n)$ can be easily computed from the recurrence (\ref{eq:pascal})~\cite{fn5}.

Finally, for the case of two coupled, correlated binomial distributions with
probabilities $p_A$ for ``success A'', $p_B$ for ``success B'' and $(1-p_A-p_B)$ for
``failure'', similar considerations lead to a recurrence relation
\begin{eqnarray}
  \label{eq:pascal2}
  P_N(n_A,n_B) & = & [1-p_A(n_A,n_B)-p_B(n_A,n_B)] P_{N-1}(n_A,n_B) + \nonumber \\
  & & p_A(n_A-1,n_B)P_{N-1}(n_A-1,n_B) +\nonumber \\
  & & p_B(n_A,n_B-1)P_{N-1}(n_A,n_B-1),
\end{eqnarray}
from which the distributions $P_N(n_A,n_B)$ for the model variants ``A'', ``B'' and
``C'' can be easily computed in O($N^3$) time.

\newpage
\begin{table}[h]
\caption{\label{tab_ddr_dfb}Fits of the phenomenological distributions
  (\ref{eq:poisson}), (\ref{eq:neg_binomi}) and (\ref{eq:gev}) to the data
  for the East German ``Oberliga'' between 1949/50 and 1990/91 and for the West
  German ``Bundesliga'' for the seasons of 1963/64 -- 1990/91.}
    \begin{tabular}{llrrrr}
         \hline \hline
         &&\multicolumn{2}{c}{Oberliga}&\multicolumn{2}{c}{Bundesliga}\\ 
       \makebox[1.5cm][c]{}&\makebox[1cm][c]{}&\makebox[2cm][c]{Home}&\makebox[2cm][c]{Away}
                                            &\makebox[2cm][c]{Home}&\makebox[2cm][c]{Away}\\ \hline
Poisson  &$\lambda$    & $ 1.85 \pm 0.02 $   & $1.05 \pm 0.01$ & $ 2.01 \pm 0.02 $  & $1.17 \pm 0.01$  \\ \hline
         &$\chi^2/{\rm d.o.f.}$ & $ 12.5  $  & $ 12.8$         & $ 6.53  $   & $ 7.31$                  \\ \hline \hline
NBD      &$p$          & $ 0.17 \pm 0.01 $  & $0.14 \pm 0.01$  & $ 0.11 \pm 0.01 $  & $0.10 \pm 0.01$  \\ 
         &$r$          & $ 9.06 \pm 0.88 $  & $6.90 \pm 0.84$  & $ 15.9 \pm 2.10 $  & $11.3 \pm 1.84$  \\ \hline
         &$p_0$         & $ 0.0191$          & $0.0112$         & $0.0213$           & $0.0126$ \\  
         &$\kappa$     & $ 0.0021$          & $0.0016$         & $0.0013$           & $0.0011$ \\ \hline  
         &$\chi^2/{\rm d.o.f.}$ & $ 0.99  $  & $ 4.09$         & $ 0.68 $  & $2.29$           \\ \hline \hline
GEV      &$\xi$        & $ -0.05\pm 0.01 $  & $0.02 \pm 0.01$  & $ -0.09\pm 0.01 $  & $-0.01\pm 0.01$  \\ 
         &$\mu$        & $ 1.12 \pm 0.02 $  & $0.49 \pm 0.02$  & $ 1.28 \pm 0.02 $  & $0.58 \pm 0.02$  \\ 
         &$\sigma$     & $ 1.30 \pm 0.02 $  & $0.90 \pm 0.02$  & $ 1.36 \pm 0.02 $  & $0.96 \pm 0.02$  \\ \hline
         &$\chi^2/{\rm d.o.f.}$ & $ 1.93  $  & $ 5.04 $        & $ 1.83 $  & $ 4.74$            \\ \hline \hline
Gumbel   &$\mu$        & $ 1.12 \pm 0.02 $  & $0.48 \pm 0.02$  & $ 1.28 \pm 0.02 $  & $0.59 \pm 0.01$  \\ 
         &$\sigma$     & $ 1.25 \pm 0.01 $  & $0.92 \pm 0.01$  & $ 1.25 \pm 0.01 $  & $0.95 \pm 0.01$  \\ \hline
         &$\chi^2/{\rm d.o.f.}$ & $ 4.13  $  & $ 4.65$         & $ 12.9 $  & $ 4.06$          \\ \hline \hline
    \end{tabular}
\end{table}

\begin{table}[t]
\caption{\label{tab_frauen}Fits of the phenomenological distributions
  (\ref{eq:poisson}), (\ref{eq:neg_binomi}) and (\ref{eq:gev}) to the data
  for the German men's premier league ``Bundesliga'' between 1963/64 and 2004/05 and
  for the German women's premier league ``Frauen-Bundesliga'' for the seasons of 1997/98 -- 2004/05.}
    \begin{tabular}{llrrrr}
       \hline \hline
       &&\multicolumn{2}{c}{Bundesliga}&\multicolumn{2}{c}{Frauen-Bundesliga}\\ 
       \makebox[1.5cm][c]{}&\makebox[1cm][c]{}&\makebox[2cm][c]{Home}&\makebox[2cm][c]{Away}
                                             &\makebox[2cm][c]{Home}&\makebox[2cm][c]{Away}\\ \hline 
Poisson  &$\lambda$    & $ 1.91 \pm 0.01 $  & $1.16 \pm 0.01$  & $ 1.78 \pm 0.04$   & $1.36 \pm 0.04$  \\ \hline
         &$\chi^2/{\rm d.o.f.}$ & $ 9.21 $  & $ 9.13$          & $14.6 $   & $14.4$           \\ \hline \hline
NBD      &$p$      & $ 0.11  \pm 0.01 $  & $0.09  \pm 0.01$      & $ 0.45 \pm 0.03 $  & $0.46 \pm 0.03$  \\
         &$r$      & $ 16.24 \pm 1.82 $  & $12.08 \pm 1.69$      & $ 2.38 \pm 0.24 $  & $1.97 \pm 0.22$  \\ \hline
         &$p_0$        & $ 0.0202$          & $0.0125$         & $0.0160$           & $0.0133$ \\  
         &$\kappa$     & $ 0.0012$          & $0.0010$         & $0.0067$           & $0.0068$ \\ \hline  
         &$\chi^2/{\rm d.o.f.}$ & $ 1.08           $  & $2.22$           & $ 2.32  $          & $ 1.37$   \\ \hline \hline
GEV      &$\xi$    & $ -0.10 \pm 0.01 $  & $-0.02 \pm 0.01$ & $ 0.04 \pm 0.04 $  & $0.25 \pm 0.07$  \\
         &$\mu$    & $ 1.17  \pm 0.02 $  & $0.57  \pm 0.01$ & $ 0.83 \pm 0.08 $  & $0.77 \pm 0.07$  \\
         &$\sigma$ & $ 1.33  \pm 0.01 $  & $0.96  \pm 0.01$ & $ 1.49 \pm 0.06 $  & $1.18 \pm 0.05$  \\ \hline
         &$\chi^2/{\rm d.o.f.}$ & $ 3.43           $  & $ 7.95$          & $ 3.40 $&   $ 1.55$  \\\hline \hline 
Gumbel   &$\mu$    & $ 1.18  \pm 0.01 $  & $0.58  \pm 0.01$ &  $ 0.81 \pm 0.08 $  & $0.58 \pm 0.07$   \\
         &$\sigma$ & $ 1.21  \pm 0.01 $  & $0.94  \pm 0.01$ &  $ 1.53 \pm 0.05 $  & $1.31 \pm 0.05$   \\ \hline
         &$\chi^2/{\rm d.o.f.}$ & $ 24.5          $  & $7.26$    & $ 3.17 $&   $ 4.09$        \\ \hline \hline
    \end{tabular}
\end{table}

\begin{table}[t]
\caption{\label{tab_fits}Matching of the sums and differences of goals. Fits were
  performed to the home and away score distributions only and mean-squared deviations were
  computed for the distributions of sums and differences from
  Eqs.~(\ref{eq:poisson_sum_diff}) and (\ref{eq:nbd_sum_diff}) with the thus found
  parameters $\lambda_h$ and $\lambda_a$ resp.\ $p_{\rm h}$, $r_{\rm h}$, $p_{\rm a}$ and $r_{\rm a}$.}
    \begin{tabular}{llcccc}
      \hline \hline
       \makebox[1.5cm][c]{}&\makebox[1cm][c]{}&\makebox[3.5cm][c]{Bundesliga 04/05}&\makebox[3.5cm][c]{Bundesliga 90/91}&
        \makebox[3cm][c]{Oberliga}&\makebox[3cm][c]{Women}\\ \hline \hline
 Poisson  & $\lambda_h$    & $ 1.91 \pm 0.01$       & $2.01 \pm 0.02$ & $1.85 \pm 0.02$ & $1.78 \pm 0.04$ \\
          & $\lambda_a$    & $ 1.16 \pm 0.01$       & $1.17 \pm 0.01$ & $1.05 \pm 0.01$ & $1.36 \pm 0.04$ \\ \hline
 Home     & $\chi^2_{\rm h}/{\rm d.o.f.}$ &  $9.21$ & $6.53$          & $12.5$          & $14.6$          \\  
 Away     & $\chi^2_{\rm a}/{\rm d.o.f.}$ &  $9.13$ & $7.31$          & $12.8$          & $14.4$          \\  
 Total    & $\chi^2_\Sigma/{\rm d.o.f.}$ &   $10.7$ & $15.9$          & $16.3$          & $10.4$          \\  
 Difference & $\chi^2_\Delta/{\rm d.o.f.}$ & $67.6$ & $578$           & $474$           & $20.2$          \\ \hline \hline
NBD       & $p_h$          & $ 0.11  \pm 0.01$      & $0.11 \pm 0.01$ & $0.17 \pm 0.01$ & $0.45 \pm 0.03$ \\ 
          & $r_h$          & $ 16.24 \pm 1.82$      & $15.9 \pm 2.10$ & $9.06 \pm 0.82$ & $2.38 \pm 0.24$ \\
          & $p_a$          & $ 0.09  \pm 0.01$      & $0.10 \pm 0.01$ & $0.14 \pm 0.01$ & $0.46 \pm 0.03$ \\
          & $r_a$          & $ 12.08 \pm 1.69$      & $11.3 \pm 1.84$ & $6.90 \pm 0.84$ & $1.97 \pm 0.22$ \\ \hline
 Home     & $\chi^2_{\rm h}/{\rm d.o.f.}$ &  $1.08$ & $0.68$          & $0.99$          & $2.32$          \\   
 Away     & $\chi^2_{\rm a}/{\rm d.o.f.}$ &  $2.22$ & $2.29$          & $4.09$          & $1.37$          \\ 
 Total    & $\chi^2_\Sigma/{\rm d.o.f.}$ &   $25.1$ & $17.3$          & $8.31$          & $18.9$          \\
 Difference & $\chi^2_\Delta/{\rm d.o.f.}$ & $23.9$ & $18.0$          & $7.16$          & $3.55$          \\ \hline \hline
    \end{tabular}
\end{table}

\begin{table}[t]
\caption{\label{tab_meanner}Fit results for models ``A'' and ``B''. Fits were
  performed to the score distributions of the home and away teams only and the
  resulting model estimates for the sums and differences of goals compared to the data.}
    \begin{tabular}{llcccc}
    \hline\hline
       \makebox[1.5cm][c]{}&\makebox[1cm][c]{}&\makebox[3.5cm][c]{Bundesliga 04/05}&\makebox[3.5cm][c]{Bundesliga 90/91}&
        \makebox[3cm][c]{Oberliga}&\makebox[3cm][c]{Women}\\ \hline
Model ``A'' & $p_{0,\rm h}$      & $0.0199 \pm 0.0002$ & $0.0210 \pm 0.0002$ & $0.0188 \pm 0.0002$ & $0.0159 \pm 0.0005$ \\ 
            & $\kappa_{\rm h}$ & $0.0015 \pm 0.0001$ & $0.0016 \pm 0.0002$ & $0.0024 \pm 0.0002$ & $0.0070 \pm 0.0005$ \\
            & $p_{0,\rm a}$      & $0.0125 \pm 0.0002$ & $0.0125 \pm 0.0001$ & $0.0112 \pm 0.0001$ & $0.0132 \pm 0.0004$ \\
            & $\kappa_{\rm a}$ & $0.0012 \pm 0.0001$ & $0.0013 \pm 0.0002$ & $0.0018 \pm 0.0002$ & $0.0071 \pm 0.0007$ \\ \hline
 Home       & $\chi^2_{\rm h}/{\rm d.o.f.}$ & $1.01$ & $0.68$ & $1.07$ & $2.28$ \\
 Away       & $\chi^2_{\rm a}/{\rm d.o.f.}$ & $2.31$ & $2.37$ & $4.23$ & $1.44$ \\
 Total      & $\chi^2_\Sigma/{\rm d.o.f.}$  & $16.6$ & $11.5$ & $5.33$ & $12.4$ \\
 Difference & $\chi^2_\Delta/{\rm d.o.f.}$  & $18.6$ & $14.0$ & $5.63$ & $2.86$ \\ \hline\hline
Model ``B'' & $p_{0,\rm h}$      & $0.0200 \pm 0.0002$ & $0.0211 \pm 0.0002$ & $0.0189 \pm 0.0002$ & $0.0166 \pm 0.0005$ \\
            & $\kappa_{\rm h}$ & $1.0679 \pm 0.0060$ & $1.0695 \pm 0.0072$ & $1.1115 \pm 0.0083$ & $1.3146 \pm 0.0303$ \\
            & $p_{0,\rm a}$      & $0.0125 \pm 0.0001$ & $0.0125 \pm 0.0002$ & $0.0112 \pm 0.0001$ & $0.0138 \pm 0.0004$ \\
            & $\kappa_{\rm a}$ & $1.0932 \pm 0.0106$ & $1.1015 \pm 0.0124$ & $1.1526 \pm 0.0149$ & $1.4115 \pm 0.0543$ \\ \hline
 Home       & $\chi^2_{\rm h}/{\rm d.o.f.}$ & $1.25$ & $0.71$ & $0.75$ & $3.24$ \\
 Away       & $\chi^2_{\rm a}/{\rm d.o.f.}$ & $1.96$ & $2.02$ & $3.35$ & $0.95$ \\
 Total      & $\chi^2_\Sigma/{\rm d.o.f.}$ &  $16.9$ & $11.8$ & $5.40$ & $13.5$ \\
 Difference & $\chi^2_\Delta/{\rm d.o.f.}$ &  $18.4$ & $13.8$ & $5.26$ & $2.82$ \\ \hline\hline
    \end{tabular}
\end{table}

\begin{table}[t]
\caption{\label{tab_wm}Fit results for the qualification phase of the ``FIFA World
  Cup'' series from 1930 to 2002.} 
    \begin{tabular}{llccccc}
    \hline\hline
       \makebox[1.5cm][c]{}&\makebox[1cm][c]{}&\makebox[2cm][c]{Home}&\makebox[2cm][c]{Away} \\ \hline \hline

Poisson     &$\lambda$             & $1.53 \pm 0.02$    & $0.89 \pm 0.01$ \\ \hline
            &$\chi^2/{\rm d.o.f.}$ & $18.6$             & $25.0$          \\ \hline \hline
NBD         &$p$                   & $0.37 \pm 0.02$    & $0.38 \pm 0.02$ \\
            &$r$                   & $3.04 \pm 0.21$    & $1.76 \pm 0.12$ \\ \hline
            &$p_0$                 & $0.0154$           & $0.0094$ \\
            &$\kappa$              & $0.0051$           & $0.0053$ \\\hline
            &$\chi^2/{\rm d.o.f.}$ & $2.67$             & $2.02$ \\ \hline \hline  
GEV         &$\xi$                 & $0.11 \pm 0.02$    & $0.19 \pm 0.02$     \\
            &$\mu$                 & $0.86 \pm 0.03$    & $0.36 \pm 0.03$     \\
            &$\sigma$              & $1.21 \pm 0.03$    & $0.86 \pm 0.02$     \\ \hline
            &$\chi^2/{\rm d.o.f.}$ & $0.85$             & $1.89$              \\ \hline \hline
Gumbel      &$\mu$                 & $0.80 \pm 0.03$    & $0.25 \pm 0.03$     \\
            &$\sigma$              & $1.31 \pm 0.02$    & $0.94 \pm 0.02$     \\ \hline
            &$\chi^2/{\rm d.o.f.}$ & $3.29$             & $12.9$              \\ \hline \hline
Model ``A'' &$p_0$                 & $0.0152 \pm 0.0003$& $0.0093 \pm 0.0002$ \\
            &$\kappa$              & $0.0053 \pm 0.0003$& $0.0055 \pm 0.0003$ \\ \hline
            &$\chi^2/{\rm d.o.f.}$ & $2.88$             & $2.19$              \\ \hline \hline
Model ``B'' &$p_0$                 & $0.0155 \pm 0.0002$& $0.0095 \pm 0.0002$ \\
            &$\kappa$              & $1.2780 \pm 0.0130$& $1.4775 \pm 0.0343$ \\ \hline
            &$\chi^2/{\rm d.o.f.}$ & $0.92$             & $0.80$              \\ \hline \hline

    \end{tabular}
\end{table}




\begin{thebibliography}{99}
\bibitem{fn1}
To ensure reliable error estimates, in the fits
presented below we ignored histogram bins consisting of single or isolated entries,
i.e., outliers.
\bibitem{fn2}
The available statistics for the ``Frauen-Bundesliga''
 is too small to allow a meaningful estimation of $R$.
\bibitem{fn3}
We disregarded all games played in tournaments on neutral grounds.
\end{thebibliography}
\end{document}